\documentclass[fleqn,10pt,english]{SelfArx}
\usepackage{lipsum}

\definecolor{color1}{RGB}{0,0,90} 
\definecolor{color2}{RGB}{0,20,20} 

\usepackage{hyperref} 
\hypersetup{hidelinks,colorlinks,breaklinks=true,urlcolor=color2,citecolor=color1,linkcolor=color1,bookmarksopen=false,pdftitle={Title},pdfauthor={Author}}

 \renewcommand{\vec}[1]{\mbox{\boldmath $#1$}}

 \def\gsim{\lower.4ex\hbox{$\;\buildrel >\over{\scriptstyle\sim}\;$}}
 \def\lsim{\lower.4ex\hbox{$\;\buildrel <\over{\scriptstyle\sim}\;$}}

 \def\aap{A\&A}
 \def\apj{ApJ}
 
 \def\apjs{ApJS}

 \def\mnras{MNRAS}

\JournalInfo{Astronomy Letters, 2023, Vol. 49, No.11}
\Archive{}
\PaperTitle{Origin of the Near-Surface Shear Layer of Solar Rotation
}
\Authors{L.~L.~Kitchatinov$^{1,2}$*}
\affiliation{\textit{$^1$Institute of Solar-Terrestrial Physics,
Lermontov Str. 126A, Irkutsk, 664033, Russia \\
$^2$ Pulkovo Astronomical Observatory, Pulkovskoe Sh.65, 196140,
St-Petersburg, Russia}} \affiliation{*\textbf{E-mail}:
kit@iszf.irk.ru}

\newcommand{\keywordname}{Keywords}

\Abstract{Helioseismology has revealed an increase in the rotation
rate with depth in a thin ($\sim$30\,Mm) near-surface layer. The
normalized rotational shear in this layer is independent of
latitude. This rotational state is shown to be a consequence of the
short characteristic time of near-surface convection compared to the
rotation period, and the radial anisotropy of the convective
turbulence. Analytical derivations within mean-field hydrodynamics
reproduce the observed normalized rotational shear and are in
agreement with numerical experiments on the radiative hydrodynamics
of solar convection. The near-surface shear layer is the source of
the global meridional flow important for the solar dynamo.}

\begin{document}

\flushbottom 

\maketitle 


\thispagestyle{empty} 

\section{INTRODUCTION} 
Helioseismology has revealed a rapid increase in the rotation rate
with depth just below the solar surface (Thompson et al. 1996; Schou
et al. 1998). The near-surface shear layer (NSSL) is interesting per
se and, as Brandenburg (2005) and Pipin \& Kosovichev (2011) noted,
may be important for the solar dynamo. The origin of the NSSL
remains a significant problem in the theory of differential
rotation.

Lebedinsky (1941) was probably the first to show that stellar
convection, under the influence of the Coriolis force, can
redistribute angular momentum along the stellar radius, producing
differential rotation. This phenomenon has since been widely studied
and confirmed in various astrophysical contexts. The angular
momentum fluxes discovered by Lebedinsky are commonly referred to as
the $\Lambda$-effect (R\"udiger 1989). The $\Lambda$-effect requires
anisotropy of the turbulent mixing: the angular momentum flux is
directed downwards in the radius if the mixing intensity along the
radius is greater than along the longitude. This radial anisotropy
can be a consequence of the radial direction of the buoyancy forces
driving convection.

The explanation of the NSSL in terms of mean-field hydrodynamics
seems obvious. The standard boundary conditions require that the
angular momentum flux through the outer surface is zero. The
downward flux due to the $\Lambda$-effect is then balanced by the
transport in the opposite direction by the eddy viscosity, and the
rotation rate increases with depth (Kitchatinov 2013). However, this
is not a generally accepted explanation. Hotta et al. (2015) and
Gunderson \& Battacharjee (2019) suggested that the NSSL is produced
by a large-scale meridional flow. Jha \& Choudhuri (2021) thought
that the NSSL is a consequence of the balance between the
centrifugal and baroclinic forces.

It is noteworthy that although the radial rotational shear and the
rotation rate decrease with latitude, the normalized shear
\begin{equation}
    \frac{r}{\Omega}\frac{\partial\Omega}{\partial r} \simeq -1
    \label{1}
\end{equation}
is constant with latitude (Barekat et al. 2014). An adequate theory
should explain this fact. In this paper we show that the absence of
a latitudinal dependence of the normalized rotational shear is due
to the short correlation time of the near-surface turbulence
compared to the solar rotation period. This result follows from the
general principles of the theory and does not depend on the
(necessarily approximate) NSSL derivation method, although the
normalized rotational shear value depends on the approximations of
the theory. The quasi-linear approximation of mean-field
hydrodynamics gives the shear value of Eq.\,(\ref{1}) for the
anisotropy of turbulence found in the numerical experiments of
Kitiashvili et al. (2023).

The $\Lambda$-effect and eddy viscosities for anisotropic turbulence
with a short correlation time are discussed in section 2. As we will
see, the latitudinally constant normalized rotational shear in the
NSSL in this case follows from the general structure of the Reynolds
stress tensor. In section 3, the rotational shear in the
near-surface layer is derived in the quasi-linear approximation for
turbulence anisotropic in both velocity directions and shape of
their correlation region (formulae for the $\Lambda$-effect and eddy
viscosities are given in the Appendix). Section 4 summarizes the
results and concludes.
\section{GENERAL EXPRESSION FOR THE NEAR-SURFACE SHEAR}
The effects of turbulence in mean-field hydrodynamics are accounted
for by the Reynolds stress tensor
\begin{equation}
    R_{ij} = -\rho\langle v_i v_j\rangle ,
    \label{2}
\end{equation}
where $\rho$ is the density, $\vec{v}$ is the fluctuating velocity
($\langle\vec{v}\rangle = 0$), and the angular brackets denote an
averaging. The contributions to the Reynolds stresses responsible
for the $\Lambda$-effect ($R^\Lambda_{ij}$) and the eddy viscosities
($R^\nu_{ij}$) are distinguished in the problem of differential
rotation:
\begin{equation}
    R_{ij} = R^\Lambda_{ij} + R^\nu_{ij} .
    \label{3}
\end{equation}
To derive these contributions, the turbulent velocity $\vec{v} =
\vec{u} + \vec{u}'$ is divided into the perturbation $\vec{u}'$ by
the large-scale flow and the unperturbed \lq original' part
$\vec{u}$. Then,
\begin{equation}
    R_{ij} = -\rho\langle u'_iu_j + u_iu'_j\rangle .
    \label{4}
\end{equation}
The $\Lambda$-effect arises from the perturbations $\vec{u}'$ caused
by the Coriolis force in a reference frame rotating with a local
angular velocity $\Omega$, while the eddy viscosities result from
the perturbations by the inhomogeneous flow $\vec{V}$ in this
reference frame (Kitchatinov 2005).

For the perturbation by the Coriolis force we have the estimate:
\begin{equation}
    {\vec u}' = 2\tau\Omega\ {\vec u}\times \hat{\vec z},
    \label{5}
\end{equation}
where $\tau$ is the characteristic time scale of turbulent
convection (correlation time) and $\hat{\vec{z}}$ is a unit vector
along the rotation axis. Substituting (\ref{5}) into (\ref{4}) gives
an expression for the cross-components
\begin{eqnarray}
    R^\Lambda_{r\varphi} &=& \rho\,2\tau\Omega
    \left(\langle u_r^2\rangle - \langle
    u_\varphi^2\rangle\right)\sin\theta ,
    \nonumber\\
    R^\Lambda_{\theta\varphi} &=& \rho\,2\tau\Omega
    \left(\langle u_\theta^2\rangle - \langle u_\varphi^2\rangle\right)\cos\theta = 0
    \label{6}
\end{eqnarray}
of the stress tensor responsible for the angular momentum transport.
Here we use the usual spherical coordinates ($r,\theta ,\varphi$)
and take into account that there is only radial anisotropy in
direction of the (unit) vector $\hat{\vec r} = \vec{r}/r$ for the
original turbulence:
\begin{eqnarray}
    \langle u_iu_j\rangle &=& \frac{1}{2}
    \left(\langle u^2\rangle - \langle u_r^2\rangle\right)\delta_{ij}
    \nonumber\\
    &-& \frac{1}{2}
    \left(\langle u^2\rangle - 3\langle u_r^2\rangle\right)\hat{r}_i\hat{r}_j .
    \label{7}
\end{eqnarray}

\begin{figure}[!t]
\includegraphics[width=8 truecm]{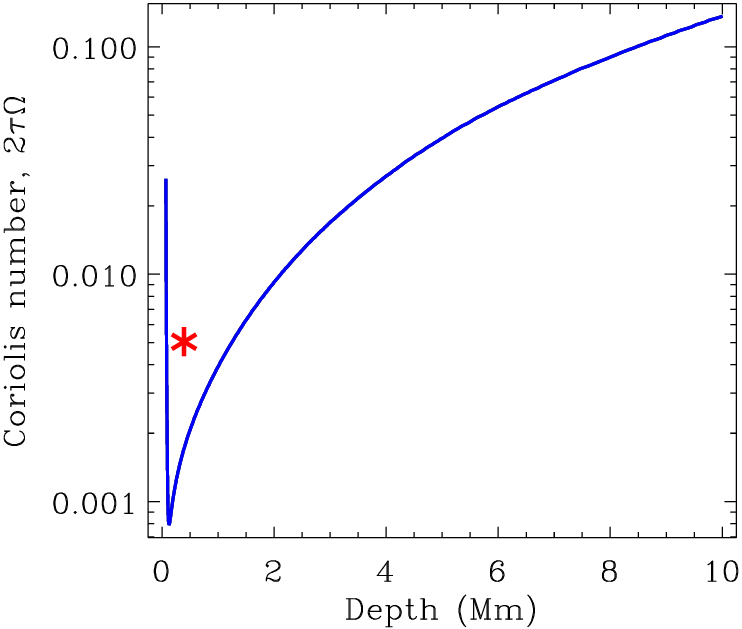}
 \caption{Coriolis number (\ref{8}) versus depth beneath the solar surface.
          The asterisk indicates the estimate for the solar granulation.}
 \label{f1}
\end{figure}

The estimations (\ref{5}) and (\ref{6}) are valid for small Coriolis
number
\begin{equation}
    \Omega^* = 2\tau\Omega ,
    \label{8}
\end{equation}
$\Omega^* \ll 1$. This number measures the intensity of the
interaction between convection and rotation. The interaction is weak
for small Coriolis number. This is why the small contribution
$\langle u_i'u_j'\rangle$ is omitted in (\ref{4}). The profile of
the Coriolis number near the solar surface is shown in
Fig.\,\ref{f1}; the profile is estimated from the
MESA\footnote{https://docs.mesastar.org} model of stellar structure
and evolution of Paxton et al. (2013) (version 496f408) applied to
the Sun. The increase in $\Omega^*$ towards the surface at small
depths is caused by the convection switch-off. MESA uses the rough
mixing length approximation. The inhomogeneities in thermal
diffusivity caused by partial ionization lead to preferential scales
of solar granulation and supergranulation (Getling et al. 2013;
Shcheritsa et al. 2018). Therefore, Fig.\,\ref{f1} also shows an
estimate for solar granulation.

The Figure covers the range of depths at which Barekat et al. (2014)
found the rotational shear (\ref{1}) in the NSSL seismologically. Of
course, the NSSL does not extend to all depths. At a depth of about
30 000 km, the radial rotational shear changes its sign (Schou et
al. 1989). As this depth is approached, the Coriolis number
increases, the rotational shear decreases, and the relation
(\ref{1}) breaks down (Komm 2022; Antia \& Basu 2022). In this paper
we consider the NSSL region where Eq.\,(\ref{1}) is valid.

As can be seen from Fig.\,\ref{f1}, the Coriolis number near the
surface is small. This greatly simplifies the theory of the
near-surface layer: the theory can be linear in rotation rate. The
estimates above have been made to demonstrate this fact. Let us now
consider the general structure of the stress tensor for small
Coriolis number.

The tensor $R^\Lambda_{ij}$ can be constructed from the angular
velocity pseudovector $\vec\Omega$, radial vector $\hat{\vec r} =
{\vec r}/r$ of anisotropy, and unitary symmetric ($\delta_{ij}$) and
fully antisymmetric ($\varepsilon_{ijk}$) tensors. For the symmetric
tensor $R^\Lambda_{ij}$ only the following structure linear in
$\vec\Omega$ is possible:
\begin{equation}
    R^\Lambda_{ij} = -\rho\nu_\Lambda\left( \hat{r}_i\varepsilon_{jkl}
    + \hat{r}_j\varepsilon_{ikl}\right)\hat{r}_k\Omega_l\ ,
    \label{9}
\end{equation}
where $\nu_\Lambda$ is the yet undetermined constant with the
dimension of viscosity and the repetition of subscripts means
summation. There is no other way to construct the non-dissipative
part of the Reynolds stresses linear in the angular velocity. For
example, the cannot be a structure
$\tilde{\nu}({\vec\Omega}\cdot\hat{\vec r})\hat{r}_i\hat{r}_j$ since
the coefficient $\tilde{\nu}$ must be a pseudoscalar, which is only
possible with the inclusion of the nonlinear contributions in
$\Omega$. A standard order-of-magnitude estimate, $\nu_\Lambda \sim
\tau u^2$, shows that Eq.\,(\ref{9}) is linear in the Coriolis
number.

In the deep convection zone, the Coriolis number is not small. The
nonlinear theory in angular velocity should be applied there, which
leads to a complication of the $\Lambda$-effect compared to
Eq.\,(\ref{9}) (R\"udiger et al. 2013).

The structure of the dissipative part $R^\nu_{ij}$ of the stress
tensor is even easier to determine. Since the large-scale velocity
$\vec{V}$ can only enter via its spatial derivative,
\begin{equation}
    R^\nu_{ij} = \rho{\cal N}_{ijkl} \frac{\partial V_k}{\partial r_l} ,
    \label{10}
\end{equation}
the viscosity tensor ${\cal N}_{ijkl}$ can be constructed from
$\hat{\vec r}$ and $\delta_{ij}$ only:
\begin{eqnarray}
    {\cal N}_{ijkl} &=& \nu_1(\delta_{ik}\delta_{jl} + \delta_{il}\delta_{jk})
    + \nu_2 \delta_{ij}\delta_{kl}
    \nonumber \\
    &+& \nu_3(\delta_{ik}\hat{r}_j\hat{r}_l + \delta_{jk}\hat{r}_i\hat{r}_l)
    \nonumber \\
    &+& \nu_4(\delta_{il}\hat{r}_j\hat{r}_k + \delta_{jl}\hat{r}_i\hat{r}_k)
    \nonumber \\
    &+& \nu_5 \delta_{ij}\hat{r}_k\hat{r}_l + \nu_6 \delta_{kl}\hat{r}_i\hat{r}_j
    +\nu_7 \hat{r}_i\hat{r}_j\hat{r}_k\hat{r}_l .
    \label{11}
\end{eqnarray}
The longitude-averaged velocity $\vec{V}$ in the rotating frame of
reference includes the differential rotation
$\delta\Omega(r,\theta)$ and the axisymmetric meridional
circulation.

The normalized rotational shear $\partial\ln(\Omega)/\partial\ln(r)$
near the surface can be expressed by the coefficients $\nu_\Lambda$
and $\nu_1,..., \nu_7$ of (\ref{9}) and (\ref{11}). Some methods of
the mean-field theory can be used to derive these coefficients. All
of the known methods are approximate. However, whatever method is
used, the calculations will result in a latitude independent
normalized rotational shear. This assertion follows from the
condition
\begin{equation}
    R_{r\varphi} = R^\Lambda_{r\varphi} + R^\nu_{r\varphi} = 0
    \label{12}
\end{equation}
at the surface of the convection zone. Note that $R_{r\varphi}$ is
the surface density of the azimuthal force acting on an area normal
to the radius $\vec{r}$. The boundary condition (12) requires that
the surface density of the external forces to be zero. Condition
(12) means that the global flow is controlled by internal processes
in the Sun, not by an external forcing.

Taking into account equations (\ref{9}) to (\ref{11}), condition
(\ref{12}) gives
\begin{equation}
    R_{r\varphi} = \rho\,\sin\theta\left(
    \nu_\Lambda\Omega + (\nu_1 + \nu_3)r\frac{\partial\Omega}{\partial r}
    \right) = 0.
    \label{13}
\end{equation}
Here, we take into account that the mean flow is axially symmetric
about the rotation axis and that the Reynolds stresses for a fixed
point ($r,\theta$) are defined in the coordinate system rotating
with a local angular velocity $\Omega(r,\theta)$. Therefore, the
velocity $\vec{V}$ in (\ref{10}) includes only the rotational shear
and meridional circulation, which, however, does not contribute to
$R_{r,\varphi}$ of Eq.\,(\ref{13}).

The equation for the near-surface normalized rotational shear,
\begin{equation}
    \frac{r}{\Omega}\frac{\partial\Omega}{\partial r} = -\frac{\nu_\Lambda}{\nu_1 + \nu_3} ,
    \label{14}
\end{equation}
common to all methods of mean-field hydrodynamics, follows from
Eq.\,(\ref{13}). In agreement with helioseismology, the normalized
rotational shear (\ref{14}) does not depend on latitude. However,
this latitude-independent constant depends on the (necessarily
approximate) method of its derivation. The normalized rotational
shear value (\ref{1}) of Barekat et al. (2014) can be reproduced
using the quasi-linear approximation of mean-field hydrodynamics.
\section{QUASI-LINER APPROXIMATION}
In the quasi-linear approximation, the disturbances $\vec{u}'$ of
the turbulent velocity $\vec{u}$ in Eq.\,(\ref{4}) can be obtained
from a linearized equation (see, e.g., R\"udiger et al. 2013). It is
convenient to use the Fourier transform,
\begin{equation}
    \hat{\vec u}({\vec k},\omega) = \frac{1}{(2\pi)^4}\int
    {\vec u}({\vec r},t)\exp( \mathrm{i}\omega t
    - \mathrm{i}{\vec k}\cdot{\vec r})\mathrm{d}^3{\vec r}\,\mathrm{d}t ,
    \label{15}
\end{equation}
which converts the differentiation operators into algebraic ones:
\begin{eqnarray}
    &&\hspace{-0.5 truecm}(\nu k^2-\mathrm{i}\omega)\hat{u}'_i({\vec k},\omega) =
    - 2\varepsilon_{inm}\hat{k}_n\hat{u}_m({\vec k},\omega)
    (\hat{\vec k}\cdot{\vec\Omega})
    \nonumber \\
    &&\hspace{-0.7 truecm}- \mathrm{i}(\delta_{in}-\hat{k}_i\hat{k}_n)k_m\int
    {\large[} \hat{u}_m({\vec k} - {\vec k}',\omega - \omega') \hat{V}_n(\vec{k}',\omega')
    \nonumber \\
    &&+\,\hat{u}_n({\vec k} - {\vec k}',\omega - \omega') \hat{V}_m(\vec{k}',\omega'){\large]}\,
    \mathrm{d}{\vec k}'\mathrm{d}\omega' .
    \label{16}
\end{eqnarray}
In this equation, $\hat{\vec k} = {\vec k}/k$ is a unit vector and
the incompressibility condition, $\vec{k}\cdot\hat{\vec u}$ = 0, has
been used to eliminate pressure.

To derive the stress tensor (\ref{4}), Eq.\,(\ref{16}) is multiplied
by $\hat{u}_j$, then averaged, inverse Fourier transformed, and the
expression with transposed indices $i$ and $j$ is added for
symmetry. The derivations are limited to the linear approximation in
the small parameter $k'/k \ll 1$ (the ratio of the correlation
length to the spatial scale of the mean velocity $\vec{V}$). The
first term on the right-hand side of Eq.\,(\ref{16}) then gives the
$\Lambda$-effect and the rest give the eddy viscosities.

To perform the derivations, we need the spectral tensor
$\hat{Q}_{ij}(k,\omega)$ for the original turbulence with radial
anisotropy
\begin{eqnarray}
&&\hspace{-0.8 truecm}\langle \hat{u}_i({\vec
k},\omega)\hat{u}_j({\vec k}_1,\omega_1)\rangle = \hat{Q}_{ij}({\vec
k},\omega)\delta({\vec k} + {\vec k}_1)\delta(\omega + \omega_1),
\nonumber \\[0.15 truecm]
&&\hspace{-0.5 truecm}\hat{Q}_{ij}({\vec
k},\omega)=\frac{E(k,\omega,\mu)}{16\pi k^2} \large[
(1+S)(\delta_{ij} - \hat{k}_i\hat{k}_j)
    \nonumber \\
&&\hspace{0.8 truecm}-\,S(\mu^2\delta_{ij} + \hat{r}_i\hat{r}_j -
\mu\,\hat{r}_i\hat{k}_j -\mu\,\hat{r}_j\hat{k}_i)\large] ,
    \label{17}
\end{eqnarray}
where $\mu = \hat{\vec r}\cdot\hat{\vec k} = \cos({\vec
r}\wedge{\vec k})$ is the cosine of the angle between the vectors
$\vec{r}$ and $\vec{k}$. Equation (\ref{17}) allows for two types of
anisotropy: anisotropy of directions and anisotropy of the shape of
the correlation region of the fluctuating velocity $\vec{u}$. The
parameter $S$ defines the directional anisotropy of the velocity.
For $S = 0$ the distribution of the velocity direction is isotropic,
but the dependence of the spectral function $E$ on $\mu$ accounts
for the anisotropy of the correlation region. For
$\partial^2E/\partial\mu^2
> 0$, the radial correlation scale is smaller than the horizontal
correlation scale (oblate convective cells). For
$\partial^2E/\partial\mu^2 < 0$, the correlation region is prolate
in radius. If $\partial E/\partial\mu = 0$, there is no shape
anisotropy, but the directional anisotropy may be present: $\langle
u^2_r\rangle > \langle u^2\rangle/3$ for $S < 0$ (radial type
anisotropy) and $\langle u^2_r\rangle < \langle u^2\rangle/3$ for $S
> 0$ (horizontal anisotropy).

Equation (\ref{17}) for the original turbulence differs from the one
used previously (Kitchatinov 2016a) by allowing for the shape
anisotropy. As before, the positive definiteness of the spectral
tensor imposes the inequality
\begin{equation}
    S \geq -1\,.
    \label{18}
\end{equation}
This implies that the turbulence cannot consist of radial flows
only. For maximum radial anisotropy ($S = -1$) there are horizontal
flows with $\langle u^2\rangle - \langle u^2_r\rangle > 0$. However,
the maximum value of S is not bounded, and (two-dimensional)
turbulence with $u_r = 0$ is possible. Note that if the fluctuation
spectrum $E(k,\omega)$ does not depend on $\mu$ (there is no shape
anisotropy), the following simple expression for the parameter $S$
in terms of the {\sl rms} velocities holds:
\begin{equation}
    S = {\langle u^2\rangle}/{\langle u^2_r\rangle} - 3.
    \label{19}
\end{equation}

The eddy viscosities and the coefficient $\nu_\Lambda$ of
Eq.\,(\ref{9}) for the $\Lambda$-effect, calculated in the
quasilinear approximation for the turbulence model (\ref{17}) are
given in the Appendix. Substitution of the transport coefficients in
Eq.\,(14) gives
\begin{equation}
    \frac{r}{\Omega}\frac{\partial\Omega}{\partial r} =
    \frac{S\int\limits_0^\infty \int\limits_0^\infty
    \frac{\nu k^2}{\nu^2k^4 + \omega^2}\int\limits_0^1
    E(k,\omega,\mu)(1-\mu^2)^2\mathrm{d}\mu\,\mathrm{d}k\mathrm{d}\omega}
    {4\int\limits_0^\infty \int\limits_0^\infty
    \frac{\nu^3 k^6}{(\nu^2k^4 + \omega^2)^2}\int\limits_0^1
    E(k,\omega,\mu)\mu^2(1-\mu^2)\mathrm{d}\mu\,\mathrm{d}k\mathrm{d}\omega} .
    \label{20}
\end{equation}
According to this equation, the normalized near-surface shear is
proportional to the anisotropy parameter $S$. The rotation rate
increases with depth for the radial anisotropy ($S < 0$), as it must
be the case (Lebedinsky 1941). Note that the rotational shear is
caused by the directional anisotropy of the turbulent velocity. The
shape anisotropy alone does not give such a shear.

Estimating the rotational shear requires a further simplification of
Eq.\,(\ref{19}). The so-called $\tau$-approximation (Brandenburg \&
Subramanian 2005) can be used for this purpose. The coefficient $\nu
k^2 - {\mathrm i}\omega$ on the left side of Eq.\,(\ref{16}) is
replaced by the inverse correlation time $1/\tau$. In the final
result (\ref{20}) of our quasilinear derivations, such a replacement
can be done by setting $\nu k^2 = \tau^{-1}$ and $\omega = 0$. Also
neglecting the shape anisotropy (the spectrum $E(k,\omega)$ does not
depend on $\mu$), we obtain a simple expression for the normalized
rotational shear:
\begin{equation}
    \frac{r}{\Omega}\frac{\partial\Omega}{\partial r} = S .
    \label{21}
\end{equation}
According to Eq.\,(\ref{18}), the shear value (\ref{1}) measured by
helioseismology is reproduced with the maximum possible radial
anisotropy. The maximum anisotropy may be a consequence of the
extreme anisotropy of the driving convection buoyancy forces which
point up or down in radius.

Note that the normalized rotational shear (\ref{21}) does not depend
on the intensity of the turbulence (provided that the Coriolis
number remains small). This is already clear from the general
expression (\ref{14}) for the shear. Only the anisotropy of
convective turbulence, not its intensity, is important for the
differential rotation in the NSSL.
\section{DISCUSSION AND CONCLUSIONS}
The proposed theory explains the NSSL in terms of the aniso\-tro\-py
of near-surface turbulent convection under the condition (\ref{12})
of zero surface density of external forces. Radial anisotropy with a
negative parameter $S$ (\ref{19}) corresponds to the observed
increase in the rotation rate with depth. In this case, the downward
non-dissipative flux of the angular momentum ($\Lambda$-effect) is
balanced by the counteracting viscous flux. For small Coriolis
number (Fig.\,\ref{f1}), the normalized rotational shear (\ref{14})
is constant with latitude, regardless of the theoretical method used
for its derivation. The quasi-linear approximation of the mean-field
theory reproduces the shear value (\ref{1}) measured by
helioseismology for the maximum radial anisotropy with $S = -1$.

This explanation is consistent with the numerical experiments of
Kitiashvili et al. (2023), whose 3D radiative hydrodynamic
simulations reproduced the NSSL. The cross component $R_{xz}$ of the
stress tensor ($R_{r\varphi}$ in the notation of this paper) in
their computations is small compared to the diagonal components, in
accordance with condition (\ref{12}) of this paper. Kitiashvili et
al. (2023) also gave the anisotropy of the simulated convective
turbulence. They used the anisotropy parameter
\begin{equation}
    A_V = 1 - 3\langle u_z^2\rangle/\langle u^2\rangle
    \label{Av}
\end{equation}
which was almost constant with the value of $A_V \simeq -0.5$ inside
the convection zone (see Fig.\,4 in Kitiashvili et al. 2023). A
negative $A_V$ means radial anisotropy. The parameter $A_V$
(\ref{Av}) can be converted into the parameter $S$ (\ref{19}) of
this paper,
\begin{equation}
    S = 3A_V/(1 -A_V).
    \label{SAv}
\end{equation}
For $A_V \simeq -0.5$ this gives $S \simeq -1$, for which our
derivations reproduce the normalized rotational shear value
(\ref{1}) detected by helioseismology.

The anisotropy of the turbulence could be inferred from the
observations of photospheric convection (see, e.g., Lida et al.
2010; Abramenko et al. 2013; Yelles\,Chaouche et al. 2020). The
observations clearly show the presence of radial flows in the
photosphere. However, since different velocity components are
measured by different methods (the Doppler measurements along the
line of sight and the motion of tracers in horizontal directions),
it is difficult to infer the anisotropy from the observations.

The importance of the NSSL for the solar dynamo is related in
particular to the fact that the global meridional flow is excited in
the surface layer. The meridional circulation most likely determines
the period of the solar cycle and it is responsible for the
equatorward migration of sunspots (see, e.g., the review by Hazra et
al. 2023). Condition (\ref{12}) disturbs the balance between the
centrifugal and baroclinic forces, and this imbalance drives the
meridional circulation (Kitchatinov 2016b).

Models of large-scale flows in the Sun and stars place the upper
boundary at a few percent of the radius below the photosphere to
avoid numerical difficulties with sharp near-surface stratification.
The boundary conditions in mean-field models set the radial
component of the large-scale flow to be zero, $V_r = 0$, but the
finite turbulent velocities $u_r$ at the upper boundary are included
via the condition (\ref{12}) for the Reynolds stresses. In this
case, it is possible to reproduce the increase in the rotation rate
with depth in about the lower third of the NSSL and the global
meridional flow (Kitchatinov \& Olemskoy 2011). A different approach
is used in the 3D numerical models of large-scale convection. Here,
the zero radial velocity, $v_r = 0$, is used as a boundary condition
without separation in its large-scale and turbulent components. This
excludes the $\Lambda$-effect of the Reynolds stress (\ref{12}).This
can lead to difficulties in reproducing the NSSL and the global
meridional circulation in 3D numerical simulations of large-scale
convection.

Our main conclusions can be formulated as follows
\begin{itemize}
\item The near-surface rotational shear in the
Sun is a consequence of the balance between the $\Lambda$-effect and
eddy viscosity for anisotropic turbulent convection.
\item The normalized rotational shear near the solar
surface does not depend on latitude because the characteristic time
of turbulent convection is here short compared to the rotation
period. As a consequence, the $\Lambda$-effect redistributes the
angular momentum in radius only, but not in the latitude.
\item The quasi-linear approximation of mean-field hydrodynamics
reproduces the rotational shear in the NSSL measured by
helioseismology with the turbulence ani\-so\-tropy value found in
the numerical radiative hydrodynamics experiments.
\end{itemize}
\section*{APPENDIX: Effective transport coefficients for anisotropic turbulence}
In this Appendix, turbulent transport coefficients for the
turbulence model of Eq.\,(\ref{17}) are given. The coefficient
$\nu_\Lambda$ of the $\Lambda$-effect (\ref{9}) reads
\begin{eqnarray}
    \nu_\Lambda &=& -\frac{S}{2}\int\limits_0^\infty \int\limits_0^\infty
    \frac{\nu k^2}{\nu^2k^4 + \omega^2}
    \nonumber\\
    &\times&\int\limits_0^1E(k,\omega,\mu)(1-\mu^2)^2
    \mathrm{d}\mu\,\mathrm{d}k\mathrm{d}\omega .
    \label{nu_lambda}
\end{eqnarray}
For the eddy viscosities of Eq.\,(\ref{11}) we find
\begin{eqnarray}
    \nu_1 &=& \frac{1}{4}\int\limits_0^\infty \int\limits_0^\infty
    \frac{\nu k^2}{\nu^2k^4 + \omega^2}\int\limits_0^1E(k,\omega,\mu)
    \nonumber \\
    &&\hspace{-0.6 truecm}
    \times\Bigg[ \mu^2(1+\mu^2) + S\mu^2(1-\mu^2)
    \nonumber \\
    &+& \frac{\nu^2k^4 - \omega^2}{\nu^2k^4 + \omega^2}
    (1+S)(1-\mu^2)^2\Bigg] \mathrm{d}\mu\,\mathrm{d}k\mathrm{d}\omega ,
    \label{nu_1}
\end{eqnarray}
\begin{eqnarray}
    \nu_2 &=& \frac{1}{4}\int\limits_0^\infty \int\limits_0^\infty
    \frac{\nu k^2}{\nu^2k^4 + \omega^2}\int\limits_0^1E(k,\omega,\mu)
    \Bigg[ 3 + \mu^4
    \nonumber \\
    &+& S(1-\mu^2)(3-\mu^2) - \frac{\nu^2k^4 - \omega^2}{\nu^2k^4 + \omega^2}
    \nonumber \\
    &&\hspace{-0.6 truecm}
    \times(1 - \mu^2)\left( 3 + \mu^2 + 3S(1-\mu^2)\right)\Bigg]
    \mathrm{d}\mu\,\mathrm{d}k\mathrm{d}\omega ,
    \label{nu_2}
\end{eqnarray}
\begin{eqnarray}
    \nu_3 &=& \frac{1}{4}\int\limits_0^\infty \int\limits_0^\infty
    \frac{\nu k^2}{\nu^2k^4 + \omega^2}\int\limits_0^1E(k,\omega,\mu)
    \Bigg[ 3\mu^2 - 5\mu^4
    \nonumber \\
    &-& S\mu^2(1-\mu^2) + \frac{\nu^2k^4 - \omega^2}{\nu^2k^4 + \omega^2}
    \nonumber \\
    &&\hspace{-0.6 truecm}
    \times(1 - \mu^2)\left( 5\mu^2 - 1 - S(1-\mu^2)\right)\Bigg]
    \mathrm{d}\mu\,\mathrm{d}k\mathrm{d}\omega ,
    \label{nu_3}
\end{eqnarray}
\begin{eqnarray}
    \nu_4 &=& \frac{1}{4}\int\limits_0^\infty \int\limits_0^\infty
    \frac{\nu k^2}{\nu^2k^4 + \omega^2}\int\limits_0^1E(k,\omega,\mu)
    \Bigg[ 1 - 5\mu^4
    \nonumber \\
    &+& S(1-\mu^2)(1-3\mu^2) - \frac{\nu^2k^4 - \omega^2}{\nu^2k^4 + \omega^2}
    \nonumber \\
    &&\hspace{-0.6 truecm}
    \times(1 - \mu^2)\left( 1 - 5 \mu^2 + S(1-\mu^2)\right)\Bigg]
    \mathrm{d}\mu\,\mathrm{d}k\mathrm{d}\omega ,
    \label{nu_4}
\end{eqnarray}
\begin{eqnarray}
    \nu_5 &=& \frac{1}{4}\int\limits_0^\infty \int\limits_0^\infty
    \frac{\nu k^2}{\nu^2k^4 + \omega^2}\int\limits_0^1E(k,\omega,\mu)
    \Bigg[ (1- \mu^2)
    \nonumber \\
    &&\times\left(5\mu^2 - 1 - S(1-\mu^2)\right)
    + \frac{\nu^2k^4 - \omega^2}{\nu^2k^4 + \omega^2}
    \nonumber \\
    &&\hspace{-1.2 truecm}
    \times\left( 3 - 6\mu^2 - 5\mu^4 + S(1-\mu^2)(3 - 7\mu^2)\right)\Bigg]
    \mathrm{d}\mu\,\mathrm{d}k\mathrm{d}\omega ,
    \label{nu_5}
\end{eqnarray}
\begin{eqnarray}
    \nu_6 &=& \frac{1}{4}\int\limits_0^\infty \int\limits_0^\infty
    \frac{\nu k^2}{\nu^2k^4 + \omega^2}\int\limits_0^1E(k,\omega,\mu)
    \Bigg[ 1 - 5\mu^4
    \nonumber\\
    &-& S(1-\mu^2)(3-\mu^2) + \frac{\nu^2k^4 - \omega^2}{\nu^2k^4 + \omega^2}
    \nonumber \\
    &&\hspace{-0.6 truecm}
    \times(1 - \mu^2)\left(5 \mu^2 - 1 + 3S(1-\mu^2)\right)\Bigg]
    \mathrm{d}\mu\,\mathrm{d}k\mathrm{d}\omega ,
    \label{nu_6}
\end{eqnarray}
\begin{eqnarray}
    \nu_7 &=& \frac{1}{2}\int\limits_0^\infty \int\limits_0^\infty
    \frac{\nu^3 k^6}{(\nu^2k^4 + \omega^2)^2}\int\limits_0^1E(k,\omega,\mu)
    \nonumber\\
    &&\hspace{-1.0 truecm}\times
    \Big[ 3 - 30\mu^2 + 35\mu^4 - S(1-6\mu^2+5\mu^4)\Big]\mathrm{d}\mu\,\mathrm{d}k\mathrm{d}\omega .
    \label{nu_7}
\end{eqnarray}

In the case of isotropic turbulence ($S \rightarrow 0$ and
$E(k,\omega)$ independent of $\mu$), the viscosities $\nu_1$ and
$\nu_2$ reduce to the known expressions (Stix et al. 1993),
\begin{eqnarray}
    \nu_1 &=& \frac{4}{15}\int\limits_0^\infty \int\limits_0^\infty
    \frac{\nu^3 k^6E(k,\omega)}{(\nu^2k^4 + \omega^2)^2}
    \mathrm{d}k\mathrm{d}\omega,
    \\
    \nu_2 &=& \frac{4}{15}\int\limits_0^\infty \int\limits_0^\infty
    \frac{\nu k^2(\nu^2k^4 + 5\omega^2)E(k,\omega)}{(\nu^2k^4 + \omega^2)^2}
    \mathrm{d}k\mathrm{d}\omega,
    \label{nu_isotropic}
    \nonumber
\end{eqnarray}
while $\nu_3, ...,\, \nu_7$ become zero, as should be the case.
\phantomsection
\section*{Funding}
This work was financially supported by the Ministry of Science and
High Education of the Russian Federation.
\phantomsection
\section*{Conflict of interest}
The author declares no conflict of interest.
\phantomsection
\section*{References}
\begin{description}
\item Abramenko,~V.\,I., Zank,~G.\,P., Dosch,~A., Yurchy\-shyn,~V.\,B., Goode,~P\,R.,
      Ahn,~K., \& Cao,~W. 2013, \apj\ {\bf 773}, 167
\item  Antia,~H.\,M., \& Basu,~S. 2022, \apj\ {\bf 924}, 19
\item  Barekat,~A., Schou,~J., \& Gizon,~L. 2014, \aap\ {\bf 570}, L12
\item  Brandenburg,~A. 2005 \apj\ {\bf 625}, 539
\item  Brandenburg,~A., \& Subramanian,~K. 2005, \aap\ {\bf 439}, 835
\newpage
\item  Getling,~A.\,V., Mazhorova,~O.\,S., \& Shcheritsa,~O.\,V. 2013, Geomagn. Aeron.
    {\bf 53}, 904
\item  Gunderson,~L.\,M., \& Battacharjee,~A. 2019, \apj\ {\bf 870}, 47
\item  Hazra,~G., Nandy,~D., Kitchatinov,~L.\,L., \& Choudhuri,~A.\,R. 2023,
       Space Sci. Rev. {\bf 219}, 39
\item  Hotta,~H., Rempel,~M., \& Yokoyama,~T. 2015, \apj\ {\bf 798}, 42
\item  Jha,~B.\,K., \& Choudhuri,~A.\,R. 2021, \mnras\ {\bf 506} 2189
\item  Kitchatinov,~L.\,L. 2005, Phys. Usp. {\bf 48}, 449
\item  Kitchatinov,~L.\,L. 2013, in: Solar and Astrophysical Dynamos and Magnetic Activity,
    IAU Symp. 294, (Eds. A.\,G.~Kosovichev, E.~de\,Gouveia\,Dal\,Pino, Y.~Yan,
    Cambridge, UK, Cambridge Univ. Press), p.399
\item  Kitchatinov,~L.\,L. 2016a, Astron. Lett. {\bf 42}, 339
\item  Kitchatinov,~L.\,L. 2016b, Geomagn. Aeron. {\bf 56}, 945
\item  Kitchatinov,~L.\,L., \& Olemskoy,~S.\,V. 2011, \mnras\ {\bf 411}, 1059
\item  Kitiashvili,~I.\,N., Kosovichev,~A.\,G., Wray,~A.\,A., Sady\-kov,~V.\,M., \&
    Guerrero,~G. 2023, \mnras\ {\bf 518}, 504
\item  Komm,~R. 2022, Front. Astron. Space Sci. {\bf 9}, 428
\item  Lebedinsky,~A.\,I. 1941, Astron Zh. {\bf 18}, 10
\item  Lida,~Y, Yokoyama,~T., \& Ichimoto,~K. 2010, \apj\ {\bf 713}, 325
\item  Paxton,~B., Cantiello,~M., Arras,~P., Bildsten,~L., Brown,~E.\,F. A.,
    Dotter,~A., Mankovich,~C., Montgomery,~M.\,H. et al. 2013, \apjs\ {\bf 208}, 4
\item  Pipin,~V.\,V., \& Kosovichev,~A.G. 2011, \apj\ {\bf 727}, L45
\item  R\"udiger,~G. 1989, Differential Rotation and Stellar Convection:
    Sun and Solar-Type Stars (Berlin, Akademie- Verlag)
\item  R\"udiger,~G., Kitchatinov,~L.\,L., \& Hollerbach,~R. 2013,
    Magnetic Processes in Astrophysics (Weinheim, Wiley-VCH)
\item Schou,~J., Antia,~H.\,M., Basu,~S., Bogart,~R.\,S., Bush,~R.\,I.,
    Chitre,~S.\,M., Christensen-Dalsgaard,~J., Di\,Mauro,~M.\,P. et al. 1998, \apj\
    {\bf 505}, 390
\item  Shcheritsa,~O.\,V., Getling,~A.\,V., \& Mazhorova,~O.\,S. 2018,
    Phys. Lett. A {\bf 382}, 639
\item Stix,~M., R\"udiger,~G. Kn\"olker,~M., \& and Grabowski,~U. 1993, \aap\
    {\bf 272}, 340
\item  Thompson,~M.\,J., Toomre,~J., Anderson,~E.\,R.,
    Antia,~H.\,M., Berthomieu,~G., Burtonclay,~D., Chitre,~S.\,M.,
    Chris\-ten\-sen-Dalsgaard,~J. et al. 1996, Science {\bf 272}, 1300
\item  Yelles Chaouche,~L., Cameron,~R.\,H., Solanki,~S.\,K., Riethm\"uller,~T.\,L.,
    Anusha,~L.\,S., Witzke,~V., Shapiro,~A.\,I., Barthol,~P. et al. 2020, \aap\
    {\bf 644}, 44
\end{description}
\end{document}